\input epsf
\documentstyle[preprint,eqsecnum,aps]{revtex}
\tightenlines

\def\INSERTCAP#1#2{\vbox{%
{\narrower\noindent%
\multiply\baselineskip by 3%
\divide\baselineskip by 4%
{\rm Table #1. }{\sl #2 \medskip}}
}}
 
\def\INSERTFIG#1#2#3{
\multiply\baselineskip
by 3 \divide\baselineskip by 4
\vbox{\bigskip \vbox{\hfil\epsfbox{#1}\hfill}%
{
\noindent%
{FIG.\ #2. }{\sl #3 \medskip}}
}\multiply\baselineskip by 4 \divide\baselineskip by 3 \hfill}%

\count255=\time\divide\count255 by 60 \xdef\hourmin{\number\count255}
  \multiply\count255 by-60\advance\count255 by\time
  \xdef\hourmin{\hourmin:\ifnum\count255<10 0\fi\the\count255}

\begin{document}

\setcounter{page}{1}

\draft
\preprint{\vbox{\hbox{JLAB-THY-98-04}
}}

\title{New Constraints on Dispersive Form Factor Parameterizations
from the Timelike Region}

\author{W. W. Buck\footnote{Electronic address: buck@jlab.org}}

\address{The Nuclear/High Energy Physics Research Center, Hampton
University, Hampton, VA 23668 \\ and \\Jefferson Lab, 12000 Jefferson
Avenue, Newport News, VA 23606}

\author{Richard F. Lebed\footnote{Electronic address: lebed@jlab.org}}

\address{Jefferson Lab, 12000 Jefferson Avenue, Newport News, VA
23606}

\bigskip
\date{February, 1998}

\maketitle
\begin{abstract}
We generalize a recent model-independent form factor parameterization
derived from rigorous dispersion relations to include constraints from
data in the timelike region.  These constraints dictate the
convergence properties of the parameterization and appear as sum rules
on the parameters.  We further develop a new parameterization that
takes into account finiteness and asymptotic conditions on the form
factor, and use it to fit to the elastic $\pi$ electromagnetic form
factor.  We find that the existing world sample of timelike data gives
only loose bounds on the form factor in the spacelike region, but
explain how the acquisition of additional timelike data or fits to
other form factors are expected to give much better results.  The same
parameterization is seen to fit spacelike data extremely well.
\end{abstract}




\narrowtext

\section{Introduction}

	Dispersion relations in field theory represent nothing more
than the application of Cauchy's theorem to Green functions, and yet
information obtained from these identities continues to provide
fascinating new insights into systems difficult or impossible to probe
with other techniques.  The natural advantage of the dispersive
approach within QCD is that its incorporation of nonperturbative
information is fully rigorous and model independent, since it works
directly with hadronic Green functions in the contour integral.
Thanks to Cauchy's theorem, the same quantity may be evaluated at a
kinematic point where perturbative QCD provides the best physical
description of the Green function.

	And yet the physical input is minimal.  Nothing has been
included from the fundamental theory except for the value of the Green
function at a single point and the existence of hadronic bound states
of given quantum numbers.  Clearly, QCD dispersion relations provide
only a bare framework of conditions that Green functions must satisfy
in order to be consistent with QCD (or any well defined field theory),
namely unitarity in the form of quark-hadron duality, crossing
symmetry, and analyticity outside of poles and cuts with locations
dictated by physical particle thresholds.  There still remains a large
space of possible hadronic Green functions, each of which satisfies
the dispersive constraints.  Nevertheless, the constraints can be
surprisingly restrictive, especially when a small amount of additional
physical input, such as data or the result of some model calculation,
is included.  It follows that constructing models manifestly
satisfying dispersive constraints is an economical way of enforcing
consistency with the basic features of QCD.

	The particular formulation of dispersion relations presented
here focuses upon obtaining bounds on both the size and shape of
electroweak form factors; this is a very old game, dating back into
the early 1960s\cite{hist}, long before the advent of QCD.  The
essential notion is the computation of the current two-point
correlator in two ways: at a point deep in the Euclidean region of
momentum transfer $t$, and as an integral over the cut generated by
on-shell hadronic states produced by the current.  Components of the
correlator are chosen so that both sides are positive definite,
implying that the neglect of contributions from some of the hadronic
states leads to a strict inequality on the hadronic amplitudes, {\it
i.e.}, form factors.

	The inequality thus obtained is expressed as a weighted
integral of squared form factors over the kinematic cut in $t$
corresponding to production of these hadrons.  Assuming only that the
branch point $t_+$ of this cut is the lowest real value of $t$ where
the Green function exhibits non-analytic behavior, analytic
continuation to the rest of the complex $t$ plane carries the
inequality to other kinematic regions.  The inclusion of perturbative
QCD Green functions in these studies was first made in
Ref.~\cite{BMR}.  Subsequently, it was observed that the most general
space of analytic functions satisfying the dispersive bounds obeys a
simple parameterization\cite{BGLApr}, which led to a number of
applications in the study of semileptonic decays, in which it is the
weak current that appears in the correlator.

	Recent applications use only data below $t_+$ to restrict
possible form factors.  Surely data from $t \geq t_+$ must provide
additional restrictions; however, for the semileptonic case this would
require data from pair production via a weak current, which is beyond
current experimental capabilities.  Instead, we focus here on
electromagnetic form factors, where data in the timelike region is
abundant.  Nevertheless, the mathematical techniques are the same.  We
show that data from the timelike region applied to the
parameterization of Ref.~\cite{BGLApr} leads to a number of new sum
rules on the parameters, which may be incorporated in an improved
parameterization.  These sum rules permit a more complete
characterization of the form factor at all values of $t$.

	As an explicit example, we consider the elastic $\pi$ form
factor and show that fits to the world sample of timelike data lead to
predictions on the spacelike behavior of the form factor, particularly
its normalization at $t=0$ and the pion charge radius.  These
predictions turn out to be very loose, but we shall see that this is
the result of gaps in the data resulting from the fact little data has
been taken near the pair production threshold $t = 4m_\pi^2$.  The
same parameterization using spacelike data yields an excellent fit, in
particular to the normalization of the form factor at $t=0$ and the
pion charge radius.

	In Sec.~II we review the derivation of the parameterization
from dispersive bounds, with an eye toward its application to
electromagnetic form factors.  Section III demonstrates the extension
of the dispersive bound to the inclusion of timelike data, and its
effect on the parameterization.  In Sec.~IV we discuss issues of
convergence of the parameterization in the timelike region, and
observations made in this inquiry lead us, in Sec.~V, to develop an
improved parameterization.  Section VI presents the result of fits to
the elastic $\pi$ form factor and discusses at length the quality of
the fits and extrapolations from timelike to spacelike regions.  We
also compare the results to direct fits to spacelike data using the
same parameterization, as well as fits from the literature.  Section
VII summarizes possible improvements and concludes.

\section{Review of the Dispersive Approach}

	Much of this discussion is patterned on that in
Ref.~\cite{BGL97} and references therein, but appears here again for
clarity in understanding what is to follow.  We begin with the QCD
two-point function of a vectorlike current $J^\mu$; in the
electromagnetic case, this is simply the conserved vector current
$\bar q \gamma^\mu q$.  The polarization tensor is defined by
\begin{equation} \label{pol}
\Pi^{\mu \nu}_J (q) = \frac{1}{q^2}(q^\mu q^\nu-q^2 g^{\mu\nu})
\Pi_J^T(q^2) + \frac{q^\mu q^\nu}{q^2} \Pi_J^L(q^2)
\equiv i \int d^4 \! x \, e^{iqx} \langle 0 |
{\rm T} J^\mu(x) J^{\dagger\nu}(0) |0 \rangle ,
\end{equation}
and in the case of a conserved current $J$, the polarization function
$\Pi^L$ vanishes.  The remaining function $\Pi^T$ is rendered finite
in QCD by making two subtractions, leading to the dispersion relation
\begin{equation} \label{chis}
\chi^T_J (q^2) \equiv \frac{1}{2} \frac{\partial^2 \Pi^T_J}
{\partial (q^2)^2} = {1\over\pi}\int_0^\infty dt \, \frac{{\rm
Im}\,\Pi_J^T (t)}{(t-q^2)^3} .
\end{equation}
The function $\chi^T_J (q^2)$ may be computed reliably in the
perturbative QCD operator product expansion for values of $q^2$ far
from the kinematic region where the current $J$ can create on-shell
hadronic states; for any light quark current and $Q^2 \equiv -q^2$,
this condition reads $Q^2 \gg \Lambda_{\rm QCD}^2$.
 
        Inserting a complete set of states $X$ into the two-point
function relates $\Pi^T_J$ to the production rate of hadrons from a
virtual photon,
\begin{equation} \label{ins}
{\rm Im}\,\Pi^T_J = \frac12 \sum_X (2 \pi)^4 
    \delta^4(q-p_X) | \langle 0| J | X \rangle |^2 \ ,
\end{equation}
where the sum is over all hadronic states $X$ with the same quantum
numbers as the current $J$, weighted by phase space.  It follows from
the dispersion relation (\ref{chis}) that the perturbatively evaluated
$\chi^T_J(q^2)$ equals the integrated production rate of $\gamma^* \to
X$ weighted with a smooth function of momentum transfer squared $t$.
Since the sum is positive semidefinite, one may restrict attention to
a subset of hadronic states to obtain a strict inequality.  In the
case discussed in this paper, $J^\mu = \frac{2}{3} \bar u \gamma^\mu u
- \frac{1}{3} \bar d \gamma^\mu d$, and we restrict to $X = \pi^+
\pi^-$.\footnote{The neutral pions do not appear here because of
charge conjugation.}  This places an upper bound on the
electromagnetic $\pi$ form factor $F(t)$ in the pair-production region
that takes the form
\begin{equation}\label{tbd}
{1\over \pi \chi^T(q^2)} \int_{t_+}^\infty dt \, {W(t)\ |F(t)|^2 \over 
(t-q^2)^3 } \leq 1 ,
\end{equation}
from the dispersion relation Eq.~(\ref{chis}).  Here $W(t)$ is a
computable function of $t$ that depends on phase space and the quantum
numbers of the particular form factor under consideration.
 
        Using analyticity to turn (\ref{tbd}) into a constraint in the
spacelike region in $t$ requires that the integrand is analytic below
the pair production threshold $t < t_+$.  To do this, we introduce a
parameter $t_s < t_+$ and a function
\begin{equation}\label{zblaschke}
z(t; t_s) = { \sqrt{t_+ -t} - \sqrt{t_+ - t_s}  \over \sqrt{t_+ -t}
+ \sqrt{t_+ - t_s} }
\end{equation}
that is real for $t < t_+$, zero at $t=t_s$, and a pure phase for $t
\ge t_+$. Any poles in the integrand of Eq.~(\ref{tbd}) can be removed
by multiplying the integrand by various powers of $z(t;t_s)$, provided
the positions $t_s$ of the sub-threshold poles in $F(t)$ are known.
Each pole has a distinct value of $t_s$, and the product $z(t;t_{s1})
z(t;t_{s2}) \cdots$ serves to remove all of them.  Such poles arise as
the contribution of resonances with masses below $\sqrt{t_+}$ to the
form factor $F(t)$, as well as singularities in the kinematic part of
the integrand.  After determining these positions from the hadronic
mass spectrum, the upper bound on $F(t)$ becomes
\begin{equation} \label{tdisp}
\frac{1}{\pi} \int_{t_+}^\infty dt \left| \frac{dz(t;t_0)}{dt} \right|
\cdot | \phi(t;t_0) P(t) F(t) |^2 \le 1 \ ,
\end{equation}
where the weight function $\phi(t; t_0)$ (known as an {\it outer
function} in complex analysis) is given by
\begin{equation}
 \phi(t;t_0) = \tilde P(t) \Biggl[ {W(t) \over 
  |dz(t;t_0)/dt| \, \chi^T(q^2) (t-q^2)^3 } \Biggr]^\frac12 \ .
\end{equation}
The factor $\tilde P(t)$ is a product of $z(t; t_s)$'s and $\sqrt{z(t;
t_s)}$'s, with $t_s$ chosen to remove the sub-threshold singularities
and cuts in the kinematic part of the integrand, while the {\it
Blaschke factor\/} $P(t)$ is a product of $z(t; t_p)$'s with $t_p$
chosen to be the positions of sub-threshold poles in $F(t)$.  The
functions $\phi(t; t_0)$ and $\tilde P (t)$ also depend on $q^2$,
which we leave implicit for notational simplicity, while $t_0$ is the
(yet to be chosen) value of $t$ for which $z(t;t_0)=0$.
 
	The dispersion inequality expressed in terms of $z$ reads
\begin{equation} \label{dispz}
\frac{1}{2\pi i} \int_C \frac{dz}{z} |\phi(z) P(z) F(z)|^2 \leq 1 ,
\end{equation}
where $F(z)$ means $F[z(t;t_0)]$ and so on.  Lacking poles, the
quantity $\phi(t;t_0) P(t) F(t)$, is expected to be analytic within
the unit disc,\footnote{Of course, subthreshold singularities due to
multiparticle or anomalous thresholds must be considered.  In many
cases\cite{BGL97,BGLAug,BL} these singularities treated as cuts are
numerically unimportant, and in the $K \pi$\cite{LS}, $\pi^+ \pi^-$,
or $K \bar K$ cases they are absent.  Moreover, we here correct an
oversight of these previous works: For on-shell multiparticle
resonances of the correlator or triangle-type anomalous threshold
diagrams, the kinematics of the internal particles is completely
fixed, meaning that the singularity in $t$ is a pole, not a cut.} and
may be expanded in a set of orthonormal functions that are simply
powers of $z(t; t_0)$.  The result of expanding in $z(t; t_0)$ is an
expression for $F(t)$ valid even in the spacelike region,
\begin{equation} \label{masterparam}
F(t) = {1 \over P(t) \phi(t;t_0) } \sum_{n=0}^\infty
                   a_n \, z(t;t_0)^n \ \ ,
\end{equation}
where, as a result of Eq.~(\ref{dispz}), the coefficients $a_n$ are
unknown constants obeying
\begin{equation} \label{abound}
\sum_{n=0}^\infty  |a_n|^2 \le 1 \ .
\end{equation}

	The functions $P(t)$, $\phi(t;t_0)$, and $z(t;t_0)$ are real
by construction for $t< t_+$.  Moreover, physical cuts in the form
factor, which lie on the real axis, generate discontinuities only in
the imaginary part of $F$.  Since there is no physical distinction
between the upper and lower complex half-planes, one must have $|F(t +
i\epsilon)| = |F(t - i\epsilon)|$ for all real $t$, and consequently
the form factor satisfies the Schwarz reflection
principle.\footnote{Alternately, the form factor is initially defined
only in the upper $t$ half plane, and the Schwarz reflection principle
continues it into the lower half plane.}  It follows from analytic
continuation away from the cut that the form factor is real on the
real $t$ axis below threshold, which maps to the real $z$ axis, and
therefore that the coefficients $a_n$ are real.

\section{Constraints from the Timelike Region}

	The constraints derived in the previous section require no
input from the value of the form factor $F(t)$ for $t \geq t_+$.
Clearly, if such information is supplied from another source such as
direct measurement, the bound provided by, {\it e.g.}, (\ref{tbd})
should be strengthened.  In terms of the equivalent bound of
Eq.~(\ref{dispz}), such data appears on a segment of the unit circle
in $z$.  Suppose that one is given data for $F(t)$ from threshold up
to some $t_u$.  According to (\ref{zblaschke}), the segment $(t_+ \mp
i \epsilon , t_u \mp i \epsilon)$ written in terms of angles from $z =
e^{i \theta}$ occupies the circular segments $\theta \in ( \theta_u ,
\pi)$ and $(-\pi, -\theta_u)$, where
\begin{equation}
\theta_u = \cos^{-1} \left[ \frac{(t_u - t_+) - (t_+ - t_0)}{t_u -
t_0} \right] .
\end{equation} 
The additional information tells us about the integrand of
Eq.~(\ref{dispz}) directly on the unit circle; here the Blaschke
factor $P(z)$ is unimodular, and so we may express this input
as follows:
\begin{eqnarray} \label{deriv}
\left| \phi(z=e^{i\theta_u}) F(z=e^{i\theta_u}) \right|^2 & = & \left|
\sum_{n=0}^\infty a_n z^n \right|^2 \nonumber \\ & = & \left(
\sum_{n=0}^\infty a_n z^n \right) \left( \sum_{m=0}^\infty a_m^* \bar
z^m \right) \nonumber \\ & = & \sum_{n=0}^\infty a_n \sum_{m=0}^\infty
a_m^* z^{n-m} \nonumber \\ & = & \sum_{n=0}^\infty a_n
\sum_{m=0}^\infty a_m^* \left[ \cos (n-m) \theta_u + i \sin (n-m)
\theta_u \right] .
\end{eqnarray}
Since the value of the modulus $|\phi(z) F(z)|$ is invariant under
reflection about the cut (which corresponds to $\theta_u \to
-\theta_u$), the left hand side of the corresponding expression for $z
= e^{-i\theta_u}$ is identical, while the right hand side flips the
sign of the sine term; this term must therefore vanish.  Equivalently,
noting that the left hand side is real and recalling from Sec.~II that
the $a_n$'s are real, the purely imaginary sine term must again
vanish.  Yet another way of seeing the same result is by using $\sin
(n-m) \theta_u = \sin n\theta_u \cos m\theta_u - \cos n\theta_u \sin
m\theta_u$ and writing the full term as
\begin{equation}
\left( \sum_{n=0}^\infty a_n \sin n\theta_u \right) \left(
\sum_{m=0}^\infty a_m^* \cos m\theta_u \right) - \left(
\sum_{n=0}^\infty a_n \cos n\theta_u \right) \left( \sum_{m=0}^\infty
a_m^* \sin m\theta_u \right) ,
\end{equation}
which is pure imaginary and vanishes trivially if all the $a_n$'s are
real.  We are therefore left with one nontrivial sum rule:
\begin{equation} \label{newsuma}
\left| \phi(z=e^{i\theta_u}) F(z=e^{i\theta_u}) \right|^2 =
\sum_{n=0}^\infty a_n \sum_{m=0}^\infty a_m \cos (n-m) \theta_u ,
\end{equation}
We arrive at the mathematical result that allows for anaylsis of
timelike data.  Its derivation has been exceptionally straightforward,
but a number of comments are in order before proceeding.

	First, the derivation is completely consistent with the
constraint (\ref{abound}).  Indeed, integrating $\cos(m-n)\theta$ over
the unit circle gives $2\pi \delta_{mn}$, and thus (\ref{newsuma})
integrates to (\ref{abound}), using (\ref{dispz}).

	Second, although the first line of (\ref{deriv}) is written as
a perfect square, exploiting the reality of the $a_n$'s means that it
is more convenient to combine the two sums as in (\ref{newsuma}).
This is because the sum rule applies to the magnitude of $F$ for
$t>t_+$.  Although phase data on above-threshold form factors exists
in some cases, typically it is data for $|F|$ above threshold that is
presented in experimental papers.  Indeed, the very nature of our
dispersion relations [see (\ref{tbd})] precludes our use of such phase
information.

	Third, we immediately see from taking particular values of
$\theta_u$ in (\ref{deriv}), or even from its first line, two very
interesting special cases:
\begin{eqnarray}
\theta_u = \pi \quad & (t = t_+): \quad & |\phi(-1) F(-1)| = \left|
\sum_{n=0}^\infty a_n (-1)^n \right| , \label{pirule} \\ \theta_u = 0
\quad & (t \to +\infty): \quad & |\phi(+1) F(+1)| = \left|
\sum_{n=0}^\infty a_n \right| . \label{zrule}
\end{eqnarray}
Clearly, conditions on $\phi$ or $F$ at these special points provide
strong constraints on the parameters.  We explore this next.

\section{Convergence of the Parameterization}

	In the last section we manipulated infinite series without
regard to the fine points of convergence.  Does the derivation of our
sum rules suffer from the possibility that some of these actions are
ill-defined?  In this section we show that this is not the case.

	The central convergence problem may be posed as follows: In
deriving (\ref{abound}), we use the analyticity of the combination
$P(z)\phi(z)F(z)$ inside the unit circle to expand it in the Taylor
series $\sum_n a_n z^n$.  Since physical information ``along the cut''
actually appears at values of momentum transfer an infinitesimal
distance from the genuine cut [$t \pm i \epsilon$], it follows that
this data is actually infinitesimally inside the unit circle in $z$,
where the geometric series $\sum_n z^n$ is still (barely) convergent.
The finiteness of analytic complex functions on a compact region such
as the disc $|z| \le 1-\epsilon$, $\epsilon > 0$, gives one sense in
which the sum $\sum_n a_n z^n$ has meaning as a finite number, through
geometric convergence.  On the other hand, for $|z|=1$ the expression
$\sum_n a_n z^n$ is defined as the value of the quantity $P(z) \phi(z)
F(z)$ along (the appropriate side of) the cut, which exists since the
only place where this combination is ill defined, the cut $t \ge t_+$,
has been mapped so that its two sides are sent to the two separated
halves of the unit circle in $z$.  These two expressions for $\sum_n
a_n z^n$ at $|z|=1$ are equal due to a theorem of Abel's: In words,
since analyticity demands that the power series $\sum_n a_n z^n$
converges for all $|z|<1$, and since its value at $|z|=1$ is defined,
the $|z| \to 1$ limit of the former equals the latter.  Therefore, the
expansion as an infinite power series in $z$ with $|z|=1$ is true as a
formal statement, but does it have meaning as a useful series with
rapidly converging partial sums when we do not invoke the technicality
that we are ``just inside'' the unit circle?

	The utility of the expansion in $z$, when continued to the
sub-threshold region for semileptonic decays, depends on its geometric
convergence, using $|z| < 1$ inside the unit circle and the
boundedness of the $a_n$'s {\it via\/} (\ref{abound}).  It follows in
that case that only the first few $a_n$'s are required to describe the
form factor over the entire semileptonic
region\cite{BGLApr,BGL97,BGLAug,BL}.  However, $|z| = 1$ on the unit
circle, and so geometric convergence fails completely.  {\it \`{A}
priori\/} it seems that an arbitrarily large number of $a_n$'s is
required to describe the form factor in the region $t>t_+$.

	Yet things are not so bleak.  We now show that the analyticity
structure plus some mild physical assumptions tells us much about the
convergence properties of the parameterization, even on the unit
circle $|z| = 1$.

	To begin with, in order to use Parseval's theorem rigorously
to prove (\ref{abound}), one must be able to exchange the order of sum
$\sum_n a_n z^n$ and integral in $z$; for this purpose, we demand the
sufficient condition that $\sum_n a_n z^n$ is uniformly convergent in
$z$ for $|z| = 1$.  Now, $|P(z) \phi(z) F(z)|$ is by physical
assumption a bounded, smooth, continuous function on the compact
region represented by either half of the unit circle.\footnote{The
functions $|\phi(z)|$ and $|P(z)| = 1$ have no singularities on
$|z|=1$, as discussed below, so our statements refer to the behavior
of $|F(z)|$.  Resonances above threshold have finite widths and
residues, so $|F(z)|$ is also smooth, and we assume that $|F(z)|$ is
finite as $t \to +\infty$, {\it i.e.}, $z \to +1$.  Finally, the
one-sided limit $z \to -1$ along the circle is assumed to exist.}
However, even with these assumptions, uniform convergence could be
spoiled by a region of $z$ where the form factor is not smooth in $z$
even though it is smooth in the original kinematic variable $t$.  The
only region fitting this description is $t \to + \infty$, which is
compressed into the finite region $z \to + 1$, $|z|=1$.  That is, a
form factor could continue to oscillate smoothly in $t$ all the way
out to infinity, but mapped to $z$ this behavior would appear to
oscillate wildly.  We therefore make the additional physical
assumption that $|F(z)|$ becomes {\em featurelessly\/} smooth for $t
\to + \infty$ in order to guarantee uniform convergence in $z$ and
thus the proof of (\ref{abound}).

	Now we may consider the full power of (\ref{abound}).  Rather
than using only the boundedness of each term as in the geometric case,
we recognize that the absolute convergence of the sum of $|a_n|^2$
implies that for sufficiently large $n$, $|a_n|$ falls off faster than
$1/\sqrt{n}$.\footnote{Of course, such statements here and below refer
to the magnitudes of $a_n$ in a statistical sense.  Individual terms
may deviate above or below the given $n$ dependence, but the overall
pattern of the terms is bounded by the stated behaviors.}  Notice
that, although this new criterion restricts the pattern of $a_n$'s,
the condition of convergence of $\sum_n a_n z^n$ is still stronger.
For, one could imagine, for example, that the true form factor has
$a_n = + 1/\sqrt{\zeta(3/2) n^{3/2}}$, which satisfies $\sum_n a_n^2 =
1$ but clearly leads to divergence of $\sum_n a_n z^n$ at $z=+1$.  To
proceed further, we need additional input.

	Up to this point, we have ignored the specific form of the
function $\phi(t;t_0)$.  In the notation of \cite{BGL97}, one begins
with
\begin{equation}
{\rm Im} \, \Pi^T \ge \frac{n_I}{K \pi} (t-t_+)^{\frac a 2}
(t-t_-)^{\frac b 2} \, t^{-c} | F(t) |^2 \theta (t-t_+) ,
\end{equation}
where $a$, $b$, $c$, and $K$ are integers specific to the form factor
under consideration, and $n_I$ is an isospin Clebsch-Gordan factor.
In the electromagnetic case for pseudoscalars, only the form factor
analogous to $f_+$ in $\bar B \to D$ appears, for which $a=3$, $b=3$,
$c=2$, and $K=48$.  The kinematic factors are simply an expression of
two-body phase space, with $t_\pm = (M \pm m)^2$ for $M \to m$ decays,
which for the elastic case simplifies to $t_+ = 4M^2$, $t_- = 0$.  The
manipulations described in Sec.~II lead to
\begin{eqnarray}\label{phioft}
\phi (t;t_0) & = & \sqrt{\frac{n_I}{48 \pi \chi^T}} \,
\left( \frac{t_+ - t}{t_+ - t_0} \right)^{\frac 1 4}
\left( \sqrt{t_+ - t} + \sqrt{t_+ - t_0} \right)
\left( t_+ - t \right)^{\frac 3 4} \nonumber \\ & &
\times \left( \sqrt{t_+ - t} + \sqrt{t_+} \right)^{-\frac 1 2}
\left( \sqrt{t_+-t} + \sqrt{t_+ + Q^2} \right)^{-3} .
\end{eqnarray}
A more convenient expression for $|\phi(t;t_0)|$ with $t>t_+$ reads
\begin{equation}
|\phi (t;t_0)| = \sqrt{\frac{n_I}{48 \pi \chi^T}} \, t^{-1/4} (t -
t_+) (t_+ - t_0)^{-1/4} \sqrt{t-t_0} \, (t + Q^2)^{-3/2} .
\end{equation}
Note in particular that $\phi \to (t-t_+)^1$ as $t \to t_+$, while
$\phi \to t^{-1/4}$ as $t \to +\infty$.  The vanishing of the first
limit indicates powers of the spatial momentum [$| {\bf p} |^3$ for
the vector form factor] of the pair created near threshold, while the
vanishing of the second limit indicates the asymptotic unitarizing
behavior of $| {\bf p} |^3 / \sqrt{t}$.  Making only the mild physical
assumptions that the form factor $F$ is not infinite at threshold and
does not grow as $t \to +\infty$ [already assumed in proving
(\ref{abound})], (\ref{pirule}) and (\ref{zrule}) give the sum rules
\begin{equation} \label{sum1}
\sum_{n=0}^\infty a_n = 0, \hspace{2em} \sum_{n=0}^\infty a_n (-1)^n =
0,
\end{equation}
or equivalently,
\begin{equation}
\sum_{n=0}^\infty a_{2n} = 0, \hspace{2em} \sum_{n=0}^\infty a_{2n+1}
= 0 .
\end{equation}

	Further improvements are possible if one considers the
particular nature of the vanishing of $\phi(t;t_0)$ as $t \to t_+$.
Since the parameters $a_n$ are defined with respect to the variable
$z$, this is most obvious if one considers $\phi$ as a function of $z$
rather than $t$, in which one sees\cite{BGL97} that $\phi(z) \propto
(1+z)^{(a+1)/2}$.  For all form factors studied in that paper, $a = 1$
or 3, which originates from the suppression of pair production by
$|{\bf p}|^a$ at threshold.  For $a=1$, not only does $\phi(z)$ vanish
at $t = t_+$, but $\phi^\prime (z=-1)$ is finite, while for $a=3$,
$\phi^\prime (z=-1)$ vanishes and $\phi^{\prime\prime} (z=-1)$ is
finite.  To be explicit, $\phi(z)$ for the vector form factor written
in $z$ assumes the form
\begin{eqnarray}
\phi(z;t_0) & = & \frac{1}{\sqrt{12 \pi t_+ \chi^T}} (1+z)^2
(1-z)^{1/2} \left( 1 - \frac{t_0}{t_+} \right)^{5/4} \left[ \sqrt{1 -
\frac{t_0}{t_+}} (1+z) + (1-z) \right]^{-1/2} \nonumber \\ & & \times
\left[ \sqrt{1 + \frac{Q^2}{t_+}} (1-z) + \sqrt{1 - \frac{t_0}{t_+}}
(1+z) \right]^{-3} .
\end{eqnarray}
Writing $\sum_n a_n z^n = P(z) \phi(z) F(z)$, we obtain
\begin{eqnarray}
\sum_{n=0}^\infty a_n (-1)^n & = & (P \phi F) (-1) , \nonumber \\
\sum_{n=0}^\infty a_n n (-1)^n & = & (P \phi F)^\prime (-1) ,
\nonumber \\ \sum_{n=0}^\infty a_n n(n-1) (-1)^n & = & (P \phi
F)^{\prime\prime} (-1) .
\end{eqnarray}
Since one may confirm that $P$, $P^\prime$, and $P^{\prime\prime}$ are
finite on the circle, making the physical assumption that the $n$th
$z$ derivative of $F(z)$ near $z=-1$ is no more singular than
$(1+z)^{-n}$ gives
\begin{eqnarray} \label{sum2}
\sum_{n=0}^\infty a_n (-1)^n & = & 0 , \nonumber \\ \sum_{n=0}^\infty
a_n n (-1)^n & = & 0 \; (a=3); \qquad \neq \infty \; (a=1), \nonumber
\\ \sum_{n=0}^\infty a_n n(n-1) (-1)^n & \ne & \infty \; (a=3).
\end{eqnarray}
The second statement improves upon the constraint of the previous sum
rule by implying $|a_n| \to 0$ faster than $1/n$, rather than
$1/\sqrt{n}$: Now we may state that $\sum_n a_n z^n$ is absolutely
convergent.  Thus one may reorder terms in the sums, as was implicit
in the proof of Eq.~(\ref{newsuma}).

	Moreover, in the case of the vector form factor ($a=3$), the
second and third expressions together tell us that in fact $a_n \to 0$
at least as fast as $1/n^2$.  This leads to a tremendous improvement
in our knowledge of convergence of $\sum_n a_n z^n$ on the
circle.\footnote{A similar treatment with powers of $(1-z)$ about
$t=+\infty$ leads to the same convergence for the scalar form factor
in processes with non-conserved currents.}  Where before we knew only
that convergence occurred, now we can quantify how fast.  The relative
error of the partial sum $\sum_{n=1}^N 1/n^2$ from its exact value of
$\zeta(2) = \pi^2/6$ is bounded by $6/(\pi^2 N)$.  We find that only 3
terms are required for a 20\% relative error, 6 for 10\%, and 60 for
1\%.  In principle, it should be very a simple matter to fit the
coarse structure of the form factor over its entire kinematic range
with very few parameters; even such a structure as the large resonant
$\rho$ peak may be accommodated by this quasi-Fourier analysis.
However, it turns out that this expectation is unfulfilled, for
reasons that we now discuss.

	The problem is that we do not know which of the $a_n$'s are
most important in this convergence.  That is, they are not necessarily
the first few $\{ a_0, a_1, \cdots , a_N \}$, as for the semileptonic
decays or more generally for any spacelike factor, where moments about
$z=0$ ($t=t_0$) dominate.  However, the constraint (\ref{abound})
tells us that, if this set is appreciable in magnitude, there is
little room for higher $a_n$'s; this leads to a rapid convergence in
$n$, and consequently, one obtains a unified picture of the form
factor spanning both spacelike and timelike regions.  The alternate
possibility is that one could have many $a_n$'s much smaller than the
saturation paradigm $1/n^2$, so that convergence is painfully slow in
$n$; in this case, the fit to timelike data voraciously demands ever
higher $a_n$'s and is relatively insensitive to the lowest
coefficients, so that the spacelike data is relatively unconstrained
by the timelike data.  A detailed fit to data is required to determine
which scenario is realized.

	We pause momentarily to review the assumptions made to obtain
appropriate convergence properties of the parameterization.
Convergence of the series $\sum_n a_n z^n$ for $t \ge t_+$ is obtained
by requiring $|F(z)|$ to be a bounded, smooth, continuous function,
while in obtaining the sum rule (\ref{abound}), we additionally assume
that $|F(z)|$ becomes {\em featurelessly\/} smooth as $t \to +
\infty$.  Absolute convergence of the series uses the particular forms
of $P$ and $\phi$, and the assumption that $F^\prime (z)$ near $z=-1$
is no more singular than $(1+z)^{-1}$. Finally, the convergence of the
series like $1/n^2$ depends on the form of $\phi(z)$ for the vector
form factor and the assumption that $F^{\prime\prime} (z)$ near $z=-1$
is no more singular than $(1+z)^{-2}$.

\section{A New Parameterization}

	After the detailed discussion of the series $\sum_n a_n z^n$
and its properties, it may seem incongruous to introduce a new
parameterization for the form factor.  Yet it is entirely appropriate
to do so, since it is more natural to incorporate the sum rules
(\ref{sum1}) and (\ref{sum2}) directly into the parameterization than
to impose them by hand.  The sort of trouble one might encounter by
using the parameterization in $a_n$'s becomes clear with reflection
upon the content of the sum rules (\ref{sum2}).  One may impose
explicit constraints on a truncated set $\{ a_0, a_1, \ldots , a_N \}$
to enforce the vanishing of the first two of these as well as the
first expression in (\ref{sum1}), for example by fixing $a_0$, $a_1$,
and $a_2$ by means of the sum rules and the values of $\{ a_3, a_4,
\ldots , a_N \}$.  But then these sum rules, which came from certain
limits of the parameterization in $z$, might be accomplished in
perverse ways.  For example, the final expression in (\ref{sum2}) may
be finite but exceptionally large, leaving $|F(t_+)|$ essentially
unbounded.
	
	The cure for such phenomena is straightforward.  One simply
factors out the appropriate behavior determined by $\phi(z)$ from the
parameterization:
\begin{equation} \label{bdef}
\sum_{n=0}^{\infty} a_n z^n \equiv (1+z)^{2} (1-z)^{1/2}
\sum_{n=0}^{\infty} b_n z^n ,
\end{equation}
defining the series $\sum_n b_n z^n$.  Since the original series
$\sum_n a_n z^n$ and also the prefactor $(1+z)^{-2} (1-z)^{-1/2}$ are
analytic in $z$ inside the unit circle, the same holds for $\sum_n b_n
z^n$.  The requirements on the singularity structure of $F(z)$ as $z
\to \pm 1$ discussed in the previous section are carried verbatim to
$\sum_n b_n z^n$ since $1/\phi(z)$ has no singularities on the unit
disc except those removed by the prefactor, while the sum rules
(\ref{sum1}), (\ref{sum2}) are automatically satisfied by the
inclusion of the $z$-dependent prefactor.  The parameters $b_n$, like
$a_n$, are real since they in particular describe the form factor in
the spacelike region, where $F(z)$ and $z$ are real.  On the other
hand, the $b_n$'s no longer satisfy any particular constraints except
that $\sum_n b_n z^n$ is analytic.\footnote{An additional constraint
occurs if a particular $t \to +\infty$ or $t \to t_+$ behavior of the
form factor is assumed.}  The extra prefactor is just that appearing
explicitly in $\phi(z)$, so one obtains the effective weight function
\begin{eqnarray} \label{phitilde}
\tilde \phi(z;t_0) & = & \phi(z) (1+z)^{-2} (1-z)^{-1/2} \nonumber \\
& = & \frac{1}{\sqrt{12 \pi t_+ \chi^T}} \left( 1 - \frac{t_0}{t_+}
\right)^{5/4} \left[ \sqrt{1 - \frac{t_0}{t_+}} (1+z) + (1-z)
\right]^{-1/2} \nonumber \\ & & \times \left[ \sqrt{1 +
\frac{Q^2}{t_+}} (1-z) + \sqrt{1 - \frac{t_0}{t_+}} (1+z) \right]^{-3}
,
\end{eqnarray}
so that
\begin{equation} \label{bfit}
|F(z)| = \frac{1}{|\tilde \phi(z)|} \cdot |\sum_n b_n z^n| .
\end{equation}
The sum rule that relates timelike data points to the
parameterization, Eq.~(\ref{newsuma}), thus assumes the parallel form
\begin{equation} \label{newsumb}
\left| \tilde \phi(z=e^{i\theta_u}) F(z=e^{i\theta_u}) \right|^2 =
\sum_{n=0}^\infty b_n \sum_{m=0}^\infty b_m \cos (n-m) \theta_u .
\end{equation}
Nevertheless, it is $|\phi(z) F(z)|$ and not $|\tilde \phi(z) F(z)|$
that appears in the dispersion integral, and so the original sum rule
(\ref{abound}) in the new basis is {\em not\/} just $\sum_n b_n^2 \le
1$, but assumes the more complicated form
\begin{equation} \label{bbound}
\sum_{n=0}^\infty b_n \sum_{m=0}^\infty b_m \, \delta (m,n) \le 1 ,
\end{equation}
where
\begin{equation} \label{bcond}
\delta (m,n) = \frac{64}{\pi} \left( 4p^2 -22p +45 \right)
\prod_{i=0}^2 \left[ (2i+1)^2 - p \right]^{-1} ,
\end{equation}
with $p \equiv 4(m-n)^2$.  In the original sum rule (\ref{abound}),
where the analogue of $\delta(m,n)$ is just $\delta_{mn}$, one clearly
requires each $|a_n| \le 1$, and every nonzero $a_n$ serves to limit
the size of succeeding terms.  However, it is not quite so simple to
determine at a glance whether the parameters $b_n$ obtained from
fitting to data or a model satisfy the original dispersion relation
(\ref{dispz}).  Moreover, without the sum rules that produced a
strongly convergent series in $a_n$, the convergence of the series in
$b_n$ follows no guaranteed pattern but still may give an adequate
fit.  We will see that this is in fact true for the $\pi$ form factor.

	An important aspect of the transformation between the
parameters $a_n$ and $b_n$ represented by (\ref{bdef}) is that both
series require only a small number of parameters to describe the form
factor in the vicinity of $z=0$; such behavior is necessary for a
useful and minimal description of the spacelike form factor around
$t=t_0$.  Specifically, one finds
\begin{eqnarray}
a_0 & = & b_0 , \nonumber \\
a_1 & = & b_1 + \frac 3 2 b_0 , \nonumber \\
a_2 & = & b_2 + \frac 3 2 b_1 - \frac 1 4 b_0 ,
\end{eqnarray}
and so forth.  The essential conundrum in obtaining a parameterization
valid in both spacelike and timelike regions is that, in the spacelike
region, or at least near $t=t_0$ where data is abundant, it cannot be
very different from the parameterizations in $a_n$ or $b_n$ discussed
above.  That is, it must have a rapidly converging Taylor series about
$z=0$.  On the other hand, it must be able to recognize strongly
localized structures in the timelike region $|z|=1$, such as the
$\rho$ resonance in the present case.  We return to this point in the
next section, once we have exhibited our empirical results.

	To summarize, the parameter basis $a_n$ admits a number of sum
rules from timelike constraints which give rise to a strongly
convergent behavior.  It also admits the very simple dispersive bound
(\ref{abound}), but may exhibit pathological behavior in fits to
timelike data.  The parameter basis $b_n$ has all of the sum rules
built in and is much more stable in timelike fits, but its dispersive
bound (\ref{bbound}) is much more complicated, and in general may
converge much more slowly.  Both bases are useful in the spacelike
region, while it requires a fit to data to determine whether either is
useful in the timelike region.

\section{The Elastic $\pi$ Form Factor}

	Our intent is to study the possibility of using the
considerable $\pi^+ \pi^-$ production data in the timelike region $t
\ge t_+$, from which the form factor $|F(t)|^2$ is extracted, to fit
to the parameters $b_n$ as determined from the analytic structure of
the QCD dispersion relation (\ref{chis}) and expressed by
(\ref{phitilde})--(\ref{bcond}).  Of particular interest is the number
$N$ of parameters necessary to give a good accounting of the data, and
what these parameters tell us about the shape of the form factor in
the spacelike region, where there exists considerable data from
$\pi^+$ elastic scattering.

\subsection{Inputs}

	Let us first consider the calculation of the current two-point
correlator, as represented by its subtracted form $\chi^T_V (Q^2)$ in
(\ref{chis}).  As noted in Sec.~II, this quantity possesses a well
defined operator product expansion in inverse powers of $Q^2$.  The
first few terms are particularly simple in the case of the light
quarks $u$ and $d$.  Neglecting subleading mass corrections,
\begin{equation} \label{ope}
\chi^T_V (Q^2)^{\vphantom\dagger}_{m=0} = \frac{1}{8\pi^2 Q^2} \left(
1 + \frac{\alpha_s}{\pi} \right) - \frac{1}{12 Q^6} \left<
\frac{\alpha_s}{\pi} G_{\mu\nu}^a G^{a \, \mu\nu} \right> -
\frac{2}{Q^6} \langle m \bar q q \rangle + O \left( \frac{1}{Q^8}
\right) +O \left(\frac{\alpha_s^2}{\pi^2} \right) .
\end{equation}
Note that the $\bar q q$ condensate has been included, despite the
vanishing of the quark mass $m$, to indicate its place in the
expansion.  The electromagnetic current bilinear $J^\mu = \frac{2}{3}
\bar u \gamma^\mu u - \frac{1}{3} \bar d \gamma^\mu d$ induces an
extra factor of $(+2/3)^2+(-1/3)^2 = 5/9$ in $\chi^T_V$ through the
quark charges.  Our goal is to choose $Q^2$ as small as possible in
order to maximize the stringency of the dispersive bound through
perturbative QCD input, but large enough that the perturbative
expansion remains valid.

	Although numerical estimates for the first few condensates
certainly exist in the literature, we use them principally to
establish that region in $Q^2$ for which convergence of the expansion
is satisfactory and to obtain a numerical uncertainty on the
lowest-order result.  To be specific, we use $\langle \alpha_s G^2 /
\pi \rangle$ = 0.02--0.06 GeV$^4$ and the expression for $\alpha_s
(Q^2) / \pi$ from the three-loop beta function with $n_f = 3$ and
$\Lambda_{\rm {\overline MS}}^{n_f=3} = 380 \pm 60$ MeV.  Values for
the $\alpha_s$ correction and the gluon condensate relative to leading
order are presented in Table I for various values of $Q^2$.  We learn
that corrections become quite large for $Q^2 < 2$ GeV$^2$.  Although
the $O(\alpha_s^1)$ correction and the gluon condensate have opposite
signs, large uncertainties on the latter prevent one from knowing how
complete this cancellation might be.  Moreover, even with a
realistically small but finite quark mass, the $\bar q q$ condensate
becomes of relative size $\sim 10$\% by $Q^2 = 1$ GeV$^2$.  We
therefore conservatively choose $Q^2 = 2$ GeV$^2$, estimating
corrections to the lowest order result to be no more than 15\%.  We
thus obtain $\chi^T_V (2.0 \, {\rm GeV}^2) = (3.52 \pm 0.53) \cdot
10^{-3} \, {\rm GeV}^{-2}$.

\bigskip
\vbox{\medskip
\hfil\vbox{\offinterlineskip
\hrule

\halign{&\vrule#&\strut $\, \,$ \hfil$#$ $\, \,$ \hfil\cr
height0pt&\omit&&\omit&&\omit&\cr
& Q^2 \, ({\rm GeV}^2)  && \alpha_s (Q^2) / \pi && \left< \alpha_s G^2
/ \pi \right> & \cr
\noalign{\hrule}
& 4.0 && +0.101 \pm 0.013 && - (0.008 - 0.025) &\cr
& 3.0 && +0.109 \pm 0.015 && - (0.015 - 0.044) &\cr
& 2.0 && +0.126 \pm 0.020 && - (0.033 - 0.099) &\cr
& 1.0 && +0.175 \pm 0.037 && - (0.132 - 0.396) &\cr} \hrule} \hfil
\medskip
\\
\INSERTCAP{1}{Values of corrections to $\chi^T_V (Q^2)$ as
appearing in Eq.~(\ref{ope}) relative to leading order.  Coefficients
have been suppressed for simplicity.}}

	The isospin factor $n_I$ in Eq.~(\ref{phioft}) for the pion
case is set to unity; in previous cases, more than one isospin channel
could couple to the current $J^\mu$ in the same dispersion relation so
that $n_I >1$, thus strengthening the expression (\ref{masterparam}).
In the $\pi$ case this is no longer possible.  As we have pointed out,
$\pi^0$ pairs cannot couple to a vector current due to charge
conjugation.  Moreover the electromagnetic current possesses both $I =
0$ and 1 amplitudes, so an improvement through $n_I$ with other
contributions to the dispersive bound would first require
disentangling these amplitudes.

	The dispersive bounds may also be strengthened by the
inclusion of perturbative QCD information along the cut in the deep
Minkowski region $t \to +\infty$, as in \cite{Dono}.  However, we opt
not to do so in this work, for in this approach one must not only
select a point $t=t_*$ for the onset of this region, but employ the
perturbative expression over the whole interval $(t_*,+\infty)$ for
inclusion in the dispersion integral.  The uncertainties are therefore
those of the integrated perturbative result, rather than those of just
one point in the Euclidean case.  For our purposes here, we prefer the
more minimal approach of using only deep Euclidean QCD calculations.

	We select the parameter $t_0 = 0$.  This natural choice means
that $z=0$ occurs where current conservation normalizes the form
factor $F(t=0)$ to unity, the charge of the pion.  It further means
that the form factor near this point is well determined by the first
few $b_n$'s, owing to the geometric convergence of $\sum_n b_n z^n$.
Of course, such convergence is contingent upon the small size of
$|b_n|$, which is not so obvious as that of the $|a_n|$ [compare
(\ref{abound}) and (\ref{bbound})] but is empirically true in our
fits.

	Explicitly, the normalization of the form factor and the pion
charge radius $\langle r^2 \rangle \equiv 6 \left. ( \partial F /
\partial q^2 )\right|_{q^2=0}$ are given by
\begin{eqnarray} \label{slike}
|F(t=0)| & = & \frac{|b_0|}{|\tilde \phi(z=0)|} , \nonumber \\ \langle
r^2 \rangle & = & \frac{3}{2m_\pi} \sqrt{6 \pi \chi^T} \left(
\sqrt{1+Q^2/4m_\pi^2} + 1 \right)^3 \left[ 3 \frac{\left(
\sqrt{1+Q^2/4 m_\pi^2} - 1 \right)}{\left( \sqrt{1+Q^2/4 m_\pi^2} + 1
\right)} b_0 - b_1 \right] .
\end{eqnarray}

	The experimental data for the form factor in the timelike
region is collected from a number of sources\cite{expt}, and each data
point is regarded as having significance entirely determined by its
stated uncertainty.  This set contains many dozens of points
stretching from $q^2 = 0.1$ to almost 10 GeV$^2$ and covers decades of
experiment, although the most recent published measurements are
already over ten years old.\footnote{Recent spacelike results from
E93-021 at Jefferson Lab are not yet in print and thus have not
included in the analysis.}  To be precise, our sample consists of 145
timelike data points, which gives the absolute upper limit of the
number $N$ of parameters $b_n$ one may hope to extract from the data;
more on this in a moment.

\subsection{The Fit}

	We minimize $\chi^2$ with regard to the unknown parameters
$b_n$ subject to the sum rule $(\ref{newsumb})$ by using the
Levenberg-Marquardt method\cite{numrec}.  Naturally, the topology
described by a function with scores of parameters utterly escapes
intuitive expectations about where the minimum might lie, so several
choices of parameters for the initial iteration are selected to test
the proper convergence of the algorithm to the global minimum.
Typically, on each given run the program finds the same $\chi^2_{\rm
min}$ to within a few tenths of a percent.  Uncertainties on the
parameters are estimated by means of the covariance matrix, despite
the fact that (\ref{newsumb}) is quadratic rather than linear in the
parameters $b_n$.  Indeed, the two sets $\{ \pm b_n \}$ trivially give
the same fit, but we select $b_0 > 0$ to guarantee the positivity of
the form factor at $t = t_0 =0$.  The program is permitted to wander
freely in $\{ b_n \}$ parameter space without imposing the constraint
(\ref{newsumb}); the degree of saturation of this bound, which
expresses quark-hadron duality, is computed at the end of the fit and
is observed to satisfy the bound in all cases.  The 15\% uncertainty
in the perturbative value of $\chi^T$ appears as a systematic
uncertainty in the parameters $b_n$ since $\tilde \phi(z) \propto
(\chi^T)^{-1/2}$, so that the dispersive bound $(\ref{bbound})$ has a
linear uncertainty in $\Delta \chi^T$, meaning that the 1 on the
r.h.s. is to be replaced by 1.15.
 
	The figure of merit in these fits is $\chi^2$ per degree of
freedom (d.o.f.), where the number d.o.f.\@ is the number of data
points minus the number of fit parameters $N$.  Obviously the average
variance of the fit value of the form factor from data decreases as
$N$ is increased, but the d.o.f.\@ denominator eventually compensates
for this advantage, and so there exists a particular value of $N$ such
that $\chi^2$/d.o.f.\@ achieves a minimum.  For our sample of 145 data
points, this occurs for about 60 parameters $b_n$.

	There is another reason to choose $N$ substantially smaller
than the number of data points.  If we think of the fit as being
essentially a decomposition into Fourier modes (as evidenced by the
fact that our basis functions are $z^n = e^{in\theta}$ on the unit
circle), then extracting a number of parameters comparable to the
number of data points is analogous to probing modes with wavelengths
as small as the spacing between these points.  Spurious high-frequency
oscillations then appear in the fit, and the same phenomenon is
observed in the present case.  If one fits not to isolated data points
but a model with a continuous prediction for the form factor, then one
is free to expand to arbitrarily high harmonics; the oscillations are
an effect of finite experimental resolution.

	In fact, these spurious oscillations tend to be compounded
with even more dramatic effects when we attempt to extrapolate from
the data-rich regions around the $\rho$ peak down to $t_+$: The
oscillations assume huge proportions, many times larger than the
$\rho$ peak itself, with the few data points far below the peak
nestling themselves in the minima of these oscillations (Fig.~1).  The
cause of this particular misfortune has a very simple origin:
Analyticity in $z$ is precisely not the same as analyticity in $t$,
and the map (\ref{zblaschke}) is not conformal at the point $t=t_+$,
{\it i.e.}, $z=-1$.  This is obvious from geometric considerations:
whereas in $t$ space segments of the real axis below and above $t_+$
are parallel, the corresponding segments of the unit disk in $z$ are
the real axis and the two halves of the unit circle, which make
$90^\circ$ angles in antiparallel directions at $z=-1$.  Indeed, the
Jacobian of the map is given by
\begin{equation} \label{jac}
\frac{\partial z}{\partial t} = -\frac{(1-z)^3}{4(t_+ - t_0) (1+z)} ,
\end{equation}
meaning that {\em any\/} parameterization in the variable $z$ has
problems as $t \to t_+$.  And yet this is a necessary feature of the
dispersive analyticity constraints: It is $z$, and not $t$, that is
the natural kinematic variable for describing the analyticity of the
form factor.  Additional physical requirements or a greater density
of data are necessary in order to impose smoothness on the form factor
in the region near $t_+$.

\epsfxsize 6.0 truein
\INSERTFIG{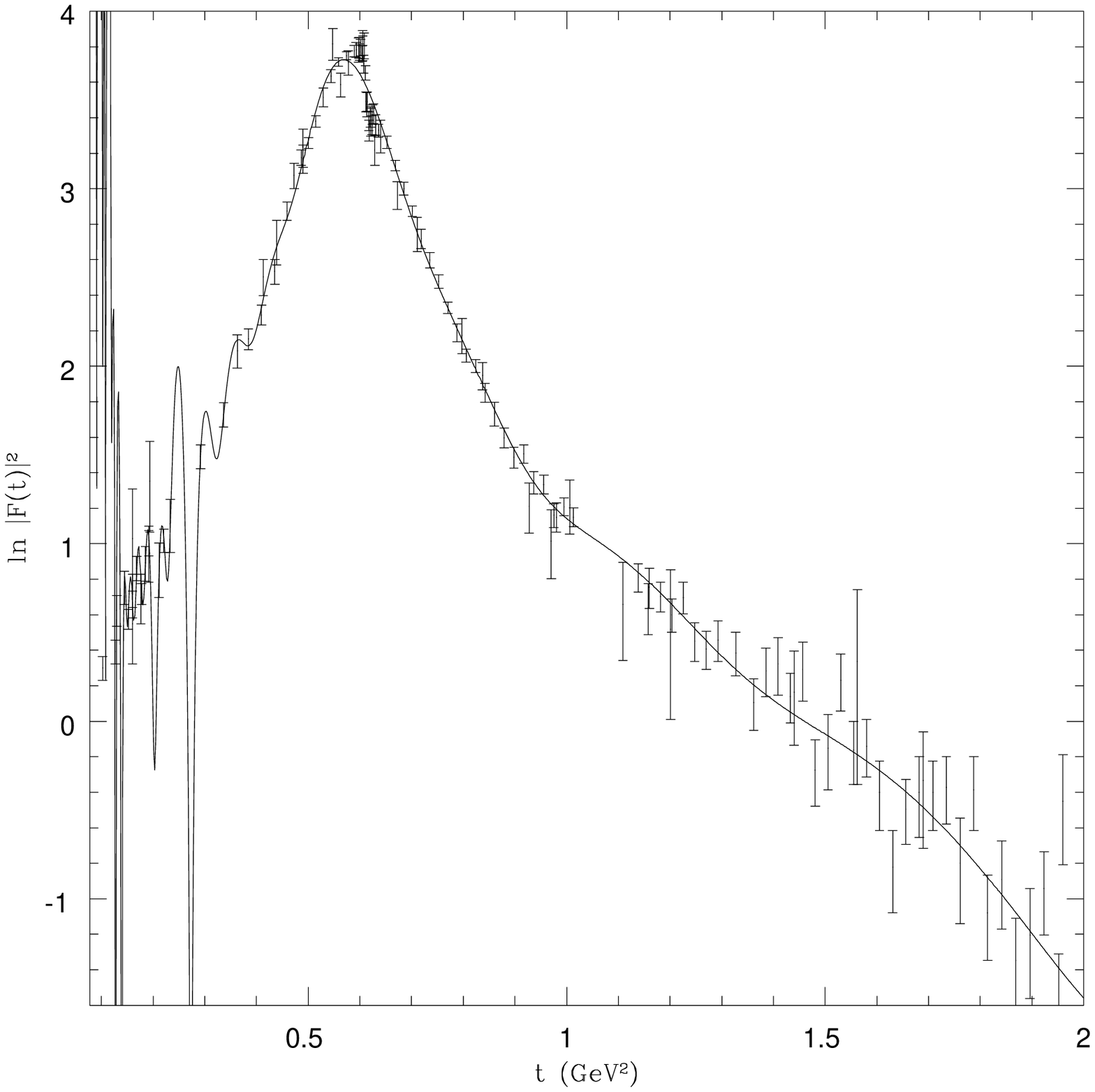}{1}{Best fit of the timelike $\pi$ elastic form
factor data $|F(t)|^2$ to the parameterization of (\ref{bfit}) with
parameters $\{ b_0, b_1, \ldots , b_{60} \}$.  The range $t_+ \le t
\le 2$ GeV$^2$ is shown, although some data exists out to 10 GeV$^2$,
in order to emphasize the $\rho$ peak.  Note especially the difficulty
of the fit in accommodating the $\omega$ shoulder, oscillations of the
fit below the peak, and the tendency of the fit to become unhinged for
$t$ near threshold where there are gaps in the data.}

\subsection{Results}

	Our best fit to the form factor over the timelike region
appears in Fig.~1.  As stated above, the fit uses $N=61$ parameters
$\{ b_0, b_1, \ldots , b_{60} \}$ and has a $\chi^2_{\rm
min}$/d.o.f.\@ of 3.20.  This large value, despite the visual goodness
of the fit in the figure, can be explained by the mild inconsistency
of data from numerous different experiments, as well as the famous
``$\omega$ shoulder'' due to the presence of destructive interference
between the narrow $\omega$ and broad $\rho$ peaks.  The shoulder
represents a sudden, nearly discontinuous change in the form factor,
which is difficult to simulate using only the lower harmonics.
Indeed, eliminating the 15 data points in the immediate vicinity of
this shoulder from the fit diminishes $\chi^2_{\rm min}$/d.o.f.\@ by
more than 1/3; of course, the oscillations in the fit below the $\rho$
peak remain.  A third possible explanation is that $\chi^2$ is large
because $N \gg 60$ might be required to fit the data adequately.
However, suppose one considers not $\chi^2_{\rm min}$/d.o.f.\@ but
rather $\chi^2_{\rm min}$/{\it datum\/} and fit to 200 parameters
(approaching the limitations of modern workstations).  Ignoring the
numerous spurious oscillations produced by this stretch, we find that
$\chi^2_{\rm min}$/{\it datum\/} decreases from 1.85 to 0.75, {\it
i.e.}, much slower than linearly with $N$.

	The degree of saturation in the best fit, namely, the fit
value for the l.h.s.\@ of (\ref{newsumb}) divided by the loosened
bound of 1.15, turns out to be 0.466, meaning that charged pions and
resonances coupled to them account for about half the dual
perturbative result at $Q^2 = 2 {\rm GeV}^2$.

	Using Eq.~(\ref{slike}), the values for the spacelike
parameters extracted from the $N=61$ fit are given by
\begin{equation}
|F(t = 0)| = 2.56 \pm 2.00, \qquad \langle r^2 \rangle = 68.2 \pm 88.3
\, {\rm GeV}^{-2} .
\end{equation}
On the surface, this is not a very impressive set of results,
considering the values extracted directly from the spacelike
data\cite{amend}:
\begin{equation}
|F(t = 0)| = 0.995 \pm 0.002, \qquad \langle r^2 \rangle = 10.26 \pm
0.26 \, {\rm GeV}^{-2} .
\end{equation}
What has gone wrong?  To sharpen this complaint, we observe that even
after fitting to a full 60 parameters, one still cannot do a very good
job extrapolating the timelike data into the spacelike region.  After
all, the naive behavior of the data plotted in $t$ seems to suggest
that the lower edge of the $\rho$ peak extrapolates smoothly into the
spacelike region to give much better figures for the normalization and
slope of the form factor at $t=0$.  However, as we have pointed out
above, the dispersive bound uses analyticity in $z$, not $t$.  Models
based upon the expected theoretical shape of the $\rho$ peak or chiral
perturbation theory implicitly assume that no peculiar behavior
afflicts the form factor near $t_+$.  For the dispersive bounds to do
the same, one would have to absorb the $(1+z)^{-1}$ factor in
(\ref{jac}) into the parameterization in order to make $\partial F/
\partial t$ finite at $t_+$.\footnote{In fact, one might expect a
discontinuity in $\partial F/\partial t$ at $t_+$, indicating the
threshold of absorptive processes entering through Im$F$.}  However,
since we have insisted upon $z$ as the natural variable of
analyticity, for the remainder of this work we maintain that such a
modification takes us too far from the original motivation of
rigorous, minimal bounds based on quark-hadron duality and
analyticity.  Certainly such modifications are straightforward to
implement, but we forgo them for now for the sake of minimality of
assumptions.

	The next issue is how one can believe the extrapolation from
the timelike to the spacelike region if indeed the best fit is
pathological in the neighborhood of $t_+$.  Again, the answer is that
it is $z$ and not $t$ that is the relevant variable of analyticity.
An extrapolation along the real $t$ axis, expressed in $z$
coordinates, consists of following the contour of the unit circle in
$z$ until reaching $t_+$ ($z=-1$), and then moving along the real $z$
axis to $t=0$ ($z=0$).  Such a path takes us straight through the eye
of the storm at $t_+$, and should be avoided if possible.  In fact,
since the form factor is analytic on the entire unit disk in $z$, one
may choose a more direct and less contentious path; for example, start
at the point on the circle where data is plentiful, such as $t_\rho
\equiv t (m_\rho^2)$ (corresponding to $\theta = 42.5^\circ$), and
proceed directly along a radius to $t=0$.  The fit for $|F|^2$ along
this contour is presented in Fig.~2.  This contour expressed in complex
$t$ space resembles a cardioid with its outer edge at $t_\rho$ and its
cusp at $t=0$.  Since it does not pass near $t_+$, it does not exhibit
strong oscillations, as is apparent from the figure.  Clearly,
analyticity in $z$ has no trouble with this region, and so the
extrapolated results have meaning.

\epsfxsize 6.0 truein
\INSERTFIG{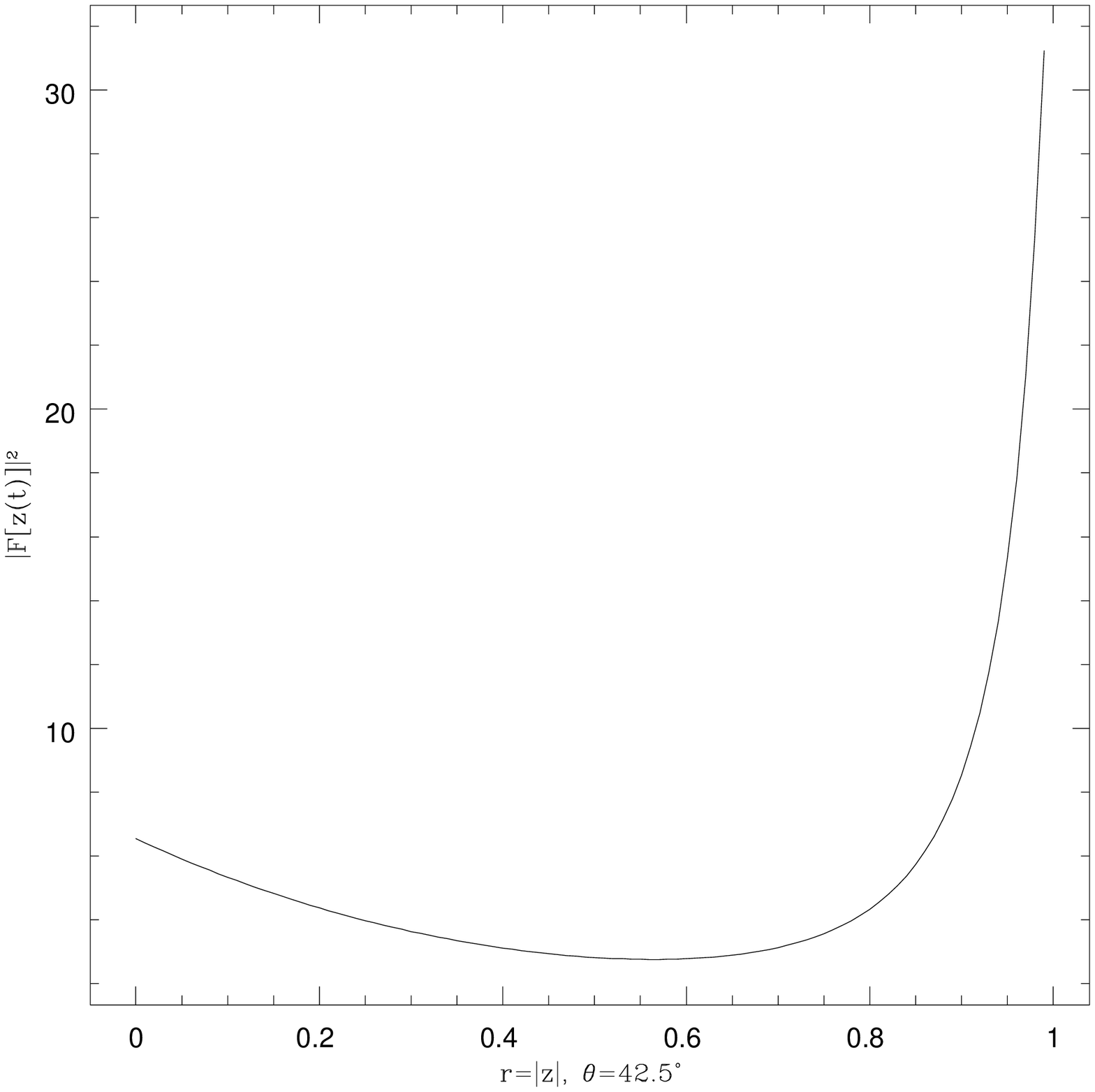}{2}{Elastic $\pi$ form factor $|F(t)|^2$ as
predicted by analyticity in $z = r e^{i\theta}$ along the radius
connecting the $\rho$ peak $t = m_{\rho}^2$ ($\theta = 42.5^\circ$) to
$t=0$ ($z=0$).  Note the absence of sudden oscillations seen in
Fig.~1.}

	So our fit values are meaningful, and the numbers extracted
for $|F(t=0)|$ and $\langle r^2 \rangle$ are certainly consistent with
those from the spacelike fit, but why are the uncertainties so large?
Again, the interpretation becomes clear in $z$ space.  The position of
points $z(t,t_0)$ in the unit disk is completely determined by the
ratios $t/t_+$ and $t_0/t_+$, as is clear from (\ref{zblaschke}).
With our choice $t_0 = 0$, the timelike data becomes compressed almost
entirely into the half-circle Re~$z > 0$: See Fig.~3.  As mentioned
above, the $\rho$ peak data is clustered about $\theta \approx \pi/4$,
while the lowest measured points are at $t \approx 0.1$ GeV$^2 \to
\theta \approx 2\pi/3$.  This is a direct result of $t_\rho \gg
(t_+ - t_0)$.  If one takes $t_0 < 0$, the extrapolation to $t_0$
should be much more precise.  Unfortunately, $t_0 = 0$ is
exceptionally convenient for the extraction of $|F(t=0)|$ and $\langle
r^2 \rangle$, as we have discussed above.  We expect these problems
would become much less severe if more near-threshold pion production
data were measured.  That this is true may be confirmed by performing
the fit after including additional Monte Carlo data points in this
region.  The extrapolation uncertainties would also decrease
substantially if we considered form factors where much more data
naturally falls just above threshold; this is precisely the situation
for the $K$ form factor, where the $\phi$ peak occurs very close to
the threshold $4m_K^2$, although in that case one faces the prospect
of depleted data for $\theta < \pi/2$.

\epsfxsize 6.0 truein
\INSERTFIG{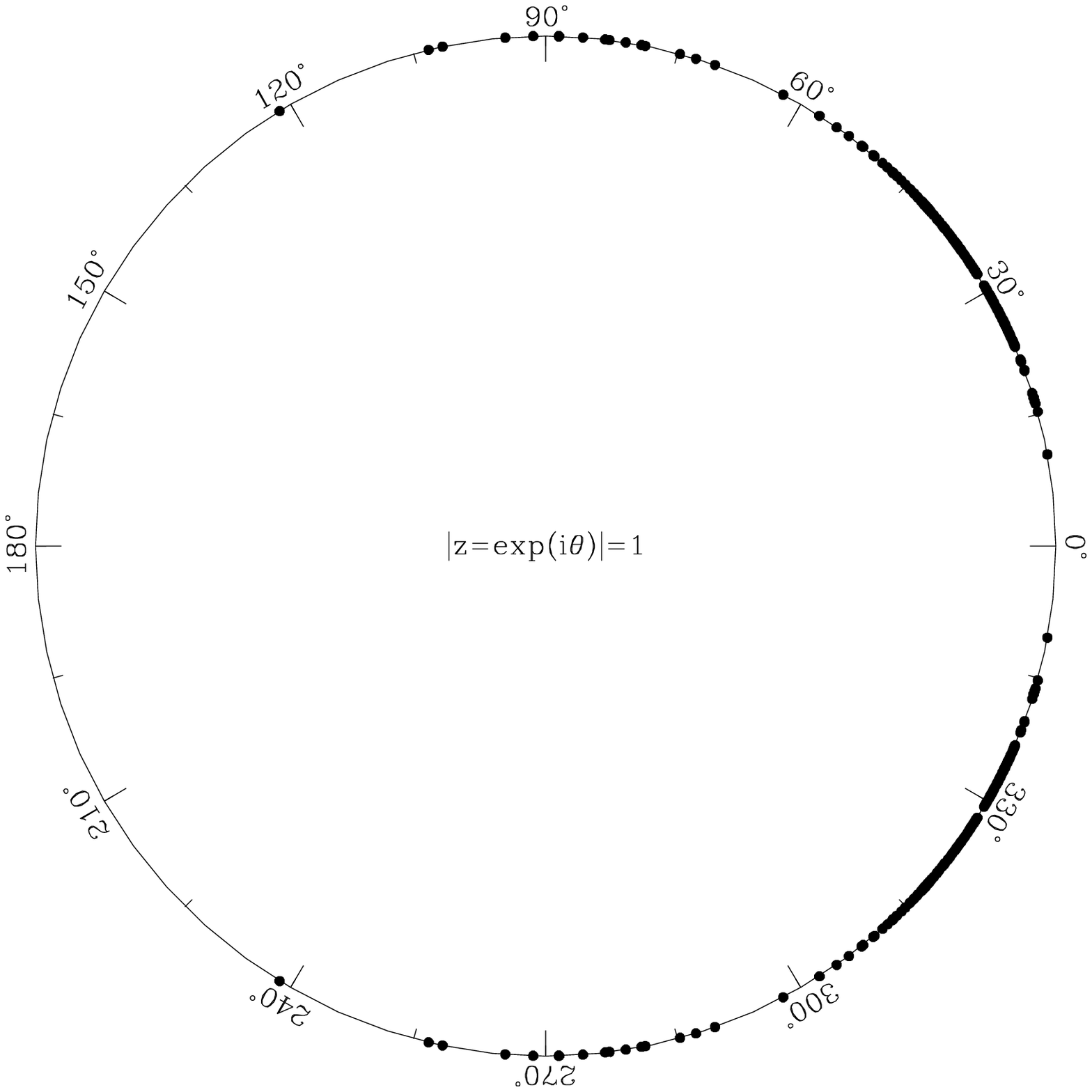}{3}{A density plot of timelike data on the unit
circle in $z$.  Note the preponderance of data near the $\rho$ peak
($\theta=42.5^\circ$) and the paucity of data near threshold $t_+$
($\theta=180^\circ$), or even beyond $\theta= 90^\circ$.}

	The distribution of the parameters $b_n$ in magnitude is an
interesting test of the convergence of the series $\sum_n b_n z^n$.
Recall that we no longer have specific mathematical information on the
rate of the convergence of this series.  In our best fit with $N=61$
we find only 3 parameters with $|b_n| > 0.05$, 12 with $0.02 < |b_n| <
0.05$, 11 with $0.01 < |b_n| < 0.02$, and 35 with $|b_n| < 0.01$.  The
largest single parameter is $b_0$, and the largest 8 parameters all
occur in $\{ b_0, \ldots , b_{10} \}$.  Although the series is not
monotonic, it appears to exhibit behavior consistent with fairly rapid
convergence in $n$.

	It is also possible to perform the fit to data fixing the
normalization $|F(t=0)| = 1$ by choosing $b_0$ {\it via\/}
Eq.~(\ref{slike}).  In a fit with $|F (t=0)| \equiv 1$ using spacelike
data, \cite{amend} gives
\begin{equation} \label{naive}
\langle r^2 \rangle =  +11.07 \pm  0.26 \, {\rm GeV}^{-2} ,
\end{equation}
whereas our fit with timelike data gives
\begin{equation}
\langle r^2 \rangle =  -13.3 \pm  27.9 \, {\rm GeV}^{-2} .
\end{equation}
Again, our fit value is consistent with the other determination.  As
expected, the uncertainty in $\langle r^2 \rangle$ decreases once
$|F(t=0)|$ is fixed, since an additional constraint has been placed on
the set of allowed analytic functions that satisfied the original
fit.  The value of $\chi^2_{\rm min}$/d.o.f.\@ for this fit is again
3.20, and the level of saturation is now 0.52.

	If enough data existed that the uncertainties on the
parameters $b_n$ were much smaller, one could extrapolate throughout
the entire spacelike region $t \in (-\infty, 0)$, which is $z \in
(0,+1)$.  The prediction for $|F|^2$ for our best fit values turns out
to fall smoothly and monotonically to zero as $t \to \infty$, and we
expect this paradigm to persist in similar fits with more constraints.
Because of the large uncertainties in our extrapolations, we decline
to exhibit just the central value fit; however, we expect this
paradigm to persist in similar fits with more restrictive constraints.

	Finally, the model-independent parameterization of
(\ref{bfit}) may be used directly in a fit to spacelike data.  This
application is entirely analogous to the semileptonic fits of
\cite{BGLApr,BGL97,BGLAug,BL,LS}, except that we pick $t_0 = 0$ and do
not here seek to minimize the ``truncation error'' of between the fit
and true form factor at points with $t \ll 0$: After all, we are
interested in moments about $t = 0$.  A collection of 45 data points
in the spacelike region taken from \cite{expt} may be fit with just
$\{ b_0, b_1, b_2 \}$ to give $\chi^2_{\rm min}$/d.o.f.\@ of only 0.97
and a saturation of the dispersion relation of only 0.018 (as
expected, fewer parameters lead to less saturation of the dispersive
bound).  From the parameter fit values we obtain
\begin{equation}
|F(t = 0)| = 0.997 \pm 0.005, \qquad \langle r^2 \rangle = 11.28 \pm
1.86 \, {\rm GeV}^{-2} ,
\end{equation}
or taking $|F(t=0)| = 1$,
\begin{equation}
\langle r^2 \rangle = 12.33 \pm 0.51 \, {\rm GeV}^{-2} ,
\end{equation}
with $\chi^2_{\rm min}$/d.o.f.\@ = 0.98 and a degree of saturation
0.022.  As a parting shot we point out that this model-independent
determination of $\langle r^2 \rangle$ differs from that obtained by a
naive linear extrapolation in $t$ (\ref{naive}) by over $2\sigma$.
The difference lies in the fact that while the usual determination
assumes $F$ linear in $t$ near $t=0$, the model-independent form
factor has a natural curvature associated with the factor $1/\tilde
\phi(z)$, which in turn arose from analyticity constraints.  The
larger size of the uncertainty in our determination takes into account
theoretical uncertainties in the shape of the form factor, which of
course are absent when one assumes linearity in $t$.

\section{Conclusions}

	We have explored some of the constraints that timelike data
and dispersive constraints place on the shape of form factors,
particularly in the spacelike region.  Much can be learned from little
more than the analytic structure of the two-point function to which
the form factor contributes and a few mild physical assumptions from
finiteness.  The sum rules and improved parameterization derived above
arise directly from these minimal considerations.

	Nevertheless, the extrapolation of a finite amount of timelike
data to the spacelike region is seen to lead to huge uncertainties in
the case of the pion elastic form factor $F(t)$.  This means that the
bare assumptions of analyticity and the applicability of QCD in the
deep Euclidean region are not enough to obtain what seems to be a
natural extrapolation of the $\rho$ peak to the spacelike region.  We
have discussed the dangers of extrapolating in the kinematic variable
$t$ rather than the natural kinematic variable of analyticity, $z$:
Namely, one must pass around a branch point $t_+ = 4m_\pi^2$ or
$z=-1$, near which the smooth behavior of the form factor may be
compromised.  In terms of $z$, the assumption of a resonant pole in
the timelike region places extremely tight constraints on the allowed
parameter space of $b_n$.

	On the other hand, the $z$ contour from timelike to spacelike
data may be distorted to avoid the troublesome point $z=-1$.  The
large uncertainties in our fit are seen to arise from the density and
placement of the data set in $z$; basically, the problem is that most
of the data, clustered near the $\rho$ peak, is far above the pair
production threshold $t_+$.  Since this is not the case for the $K$
form factor, the extrapolation may be much more reliable in that case.

	This is not to indicate that the technique is hopelessly weak
or intrinsically flawed, however.  Rather, we have endeavored to
exhibit what is obtained from the {\em absolute minimum\/} of
assumptions.  We have also suggested at various points possible
improvements to the program presented above.  It is useful to collect
them here for the convenience of the reader.  First, the calculation
of the perturbative quantity $\chi^T (Q^2)$ may be improved by a
careful analysis of multiloop effects and condensates.  We have
evaluated the perturbative result only at the single value $Q^2 = 2$
GeV$^2$.  Certainly the behavior of the dispersion relation as a
function of $Q^2$, which is equivalent to a moment analysis, provides
additional constraints.

	Second, the stringency of the dispersive bound is determined
by both $\chi^T (Q^2)$ and the level of saturation by all contributing
states excepting those being probed by the parameterization.  In the
present case, this means any states like $K \bar K$ that receive
contributions from the electromagnetic current but have not already
been counted in $\pi^+ \pi^-$ production (such as the $\rho$).  A
calculation or estimation of this contribution effectively decreases
the dispersive saturation limit allotted to the pion form factor.
Also falling into this category is the deep Minkowski region $t \gg
m_\rho^2$ of the pion form factor, for which data has not yet been
collected.  Here, one may either use its integrated weight to increase
the stringency of the dispersive bound, or retain its functional form
to obtain another sum rule on the form factor parameterization.

	The latter suggestion leads us to the third possible
improvement, that conditions on the shape of the form factor lead to
constraints on the form of the parameterization.  We have endeavored to
make as few assumptions as possible along these lines, lest the
model-independency of the parameterization is lost.  Nevertheless,
modelers are free to start with the dispersive bounds as a starting
point for building form factors that automatically satisfy nontrivial
QCD constraints.

	Another possibility along these lines not yet mentioned is the
choice of the series used to parameterize the unknown physics of the
form factor.  We have always used a simple Taylor series in $z^n$ to
express the analyticity of the form factor since the set $\{ z^n \}$
is a complete and orthonormal basis on the unit circle for analytic
functions.  Although this basis proved immensely useful (indeed,
optimal) for spacelike fits, perhaps a better basis exists for
accommodating the relatively rapid variations in the form factor due
to structure in resonant peaks, such as the $\omega$ shoulder.
However, the constraints on this new parameterization are manifold: If
it is to be adequate for both timelike and spacelike fits, it must
admit a Taylor series rapidly convergent in $z^n$ as $z \to 0$, all of
its Taylor coefficients must be real (so that the form factor is real
in the spacelike region, where $z$ is real), but it cannot converge
{\em too\/} quickly in $z^n$ as $|z| \to 1$, or else our polynomial
fits in $z^n$ would already have been adequate.  Moreover, since the
proposed functions are analytic in $z$ and not $\bar z$ on the unit
disk, natural choices such as $\cos n\theta$ and $\sin n\theta$ are
forbidden.

	Clearly, many aspects of this program yet remain to be fully
explored, but the variety of directions for improvements are quite
astounding.  Dispersive techniques possess the rare ability to
encapsulate a great deal of nontrivial physics with true simplicity
and elegance.

\vskip 0.5in {\it Acknowledgments}
\hfil\break
We are indebted to R. Lewis and R. Williams for numerous invaluable
discussions, and C. G. Boyd for comments on the manuscript.  This work
was supported by the Department of Energy under contract No.\
DE-AC05-84ER40150.  WWB also acknowledges support under NSF
Cooperative Agreement HRD-9154080.

\end{document}